\documentclass[aps,prl,twocolumn,groupedaddress]{revtex4}

\usepackage{graphicx}

\begin{document}
 
\title{X-wave mediated instability of plane waves in Kerr media}
\author{Claudio Conti}
\email{c.conti@ele.uniroma3.it}
\homepage{http://optow.ele.uniroma3.it}
\affiliation{
NooEL, Nonlinear and Optoelectronics Laboratory,
National Institute for the Physics of Matter, INFM - Roma Tre,
Via della Vasca Navale 84, 00146 Rome - Italy}
%
\date{\today}
\begin{abstract}
Plane waves in Kerr media spontaneously
generate paraxial X-waves (i.e. non-dispersive and non-diffractive pulsed beams) that get amplified along propagation.
This effect can be considered a form of conical emission
(i.e. spatio-temporal modulational instability), and can be used as 
a key for the interpretation of the out of axis energy emission in 
 the splitting process of focused pulses in normally dispersive materials. A new class of 
spatio-temporal localized wave patterns is identified.
X-waves instability, and nonlinear X-waves, are also expected
in periodical Bose condensed gases.
\end{abstract}
\maketitle
%
%
The dynamics of focused femtoseconds pulses (FFP) in optically
nonlinear media has a fundamental importance and it is relevant
in all of the applications of ultrafast optics.
The basic mechanism, when dealing with the propagation in 
normally dispersive materials, is the splitting of the pulse, 
which has been originally predicted more than ten years ago
\cite{Chernev92,Rothenberg92}
 and recently reconsidered, due to the development of the physics of FFP. 
\cite{Ranka96,Berge96,Brabec97,Trippenbach97,Ranka98,Diddams98,Zozulya99,
Litvak00,Eisenberg01,Berge02,Ward03}

If $z$ is the direction of propagation, $t$ is the time
in the reference frame where the pulse is still, and $r=\sqrt{x^2+y^2}$
 the radial cylindrical coordinate,
this process can be roughly divided into a series of steps.
They describe the propagation of a gaussian (in space and time) pulse,
which travels in a focusing medium, and undergoes relevant reshaping,
due to the interplay of diffraction, dispersion and the Kerr effect (when
the peak power is sufficiently greater than the critical power $P_c$ 
for self-focusing \cite{Chernev92,Rothenberg92}):
1) the out of axis energy is focused towards $r=0$ around $t=0$;
2) the pulse at $r=0$ is compressed;
3) lobes appear in the on-axis temporal spectrum;
4) the pulse splits in the time domain.
\\
Before the breakup,  relevant
out of axis energy emission and re-distribution occurs, as originally
described by Rothenberg in Ref. \onlinecite{Rothenberg92}. 
Looking  to the spatio-temporal profile an X-shape (or hyperbola)
 is observed before the splitting (see, e.g., figures in \cite{Litvak00,Berge02}). 
This process has been theoretically described by
different approaches \cite{Luther94,Fibich95,Litvak00,Berge02}, 
and the onset of an X-shape can be ultimately related to
the hyperbolic characteristics over which small perturbations
evolve. \cite{LeviCivita}
 This can be checked, for example, by the hydrodynamical approach
to the nonlinear Schr\"odinger equation.\cite{Staliunas93,Stringari96}

\begin{figure}
\includegraphics[width=60mm]{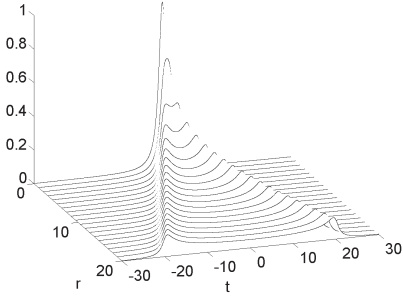}
\caption{\label{figurepsiX} Tri-dimensional plot of $\psi_X$, the simplest
radial-symmetry X-wave ($\Delta=1$).}
\end{figure}

Recently, attention has been devoted to the existence of {\it X-waves} in 
optically nonlinear media. The latter are self-trapped (i.e. non-diffracting
and non-dispersive) waves that are well known in the field of linear propagation in acustics 
\cite{Ziolkowski89,Lu92,Stepanishen97,Salo99}
and in electromagnetism
\cite{Durnin87,Recami98,Saari97,Sonajalg97,Reivelt00,Mugnai00,Sheppard01,Zamboni02b}.
The simplest X-wave has the shape of a double-cone, or clepsydra,
that appears as an X when, e.g., a section is plotted in the plane $(x,t)$,
and as V in the plane $(r,t)$, as shown in figure \ref{figurepsiX}. 
Optical X-waves in nonlinear media 
 have been theoretically predicted in \cite{Conti02} and experimental
results, with a direct observation of the conical spatio-temporal shape,
have confirmed their existence and generation in crystals for
second-harmonic generation (SH-G) \cite{QELS01,QELS02,DiTrapani03}.  
In Ref. \onlinecite{Conti02b}
 it has been theoretically  shown how, during non-depleted-pump SHG,
the phase matched spatio-temporal harmonics let the SH beam become an X-wave.
\footnote{For the sake of concreteness, I use the term ``X-wave'', without making
distinction among the many families of localized-wave patterns solutions to the
linear Helmotz equation. See,e.g., Ref. \onlinecite{Saari01}.}

Conical emission (CE), or spatio-temporal modulational instability (MI)
(\cite{Liou92,Luther94,Fonseca99,Picozzi02}),  
has been addressed in Ref. \onlinecite{Trillo02} as a basic mechanism underlaying
 the spontaneous formation of an X-wave and, as a foundation for the understaing of the splitting, in Ref. \onlinecite{Litvak00}.
In this Letter I consider the formation of X-waves 
in Kerr media (or in quadratic media, in the regime
where they mimic cubic nonlinearity, see, e.g., the chapter after Torruellas, Kivhsar and Stegeman
 in \cite{Trillo01} and Ref. \onlinecite{Buriak02}). 
A new form of instability of plane waves can be introduced
by directly involving self-localized spatio-temporal wave patterns. 
The process strongly resembles CE, i.e. the amplification of plane waves from noise,
but in this case X-waves, instead of periodical patterns, emerge from the breakup of an unstable pump beam. 
This mechanism is responsible of the first stage of
the pulse-splitting, i.e. the out-of-axis energy redistribution,
such as MI breaks a continuos wave signal
into a periodical pattern of solitons.\cite{Hasegawa84}

The wave equation describing the propagation in nonlinear Kerr media, whose refractive $n$ index obeys
 the law $n=n_0+n_2 I$, with $I$ the optical intensity, in the 
framework of the paraxial and of the slowly varying envelope approximation, is written as
\begin{equation}
\label{NLSdimensional}
i\partial_Z A+ik'\partial_T A+\frac{1}{2k}\nabla^2_{XY}A-\frac{k''}{2}\partial_{TT}A+\frac{k n_2}{n_0}|A|^2 A=0\text{,}
\end{equation}
where $A$ is normalized such that $|A|^2=I$, and I have taken $k''>0$ (i.e. the medium is normally dispersive).
$X,Y,Z,T$ are the real world variables, 
$\lambda$ the wavelength,
$n(\lambda)$ the refractive index and 
$k(\lambda)=2\pi n(\lambda)/\lambda$. Eq. (\ref{NLSdimensional})
  can be casted in the adimensional form:
\begin{equation}
\label{NLS}
i\partial_z u+\Delta_\perp u-\partial_{tt} u+\chi |u|^2 u=0\text{,}
\end{equation}
by defining $z=Z/Z_{df}$ with  
$Z_{df}=2k W_0^2$ and $W_0^2$ a reference beam waist;
$\Delta_\perp\equiv\partial_{xx}+\partial_{yy}$ the tranverse Laplacian 
with $(X,Y)=W_0 (x,y)$; $t=(T-Z/V_g)/T_0$ the retarded time 
in the frame travelling at the group velocity $V_g=1/k'$ in units of 
$T_0=(k''Z_{df}/2)^{1/2}$. The optical field envelope $A$ is given by
$A=A_0 u$, with $A_0=(n_0/k |n_2| Z_{df})^{1/2}$. $\chi>0$ ($\chi<0$) 
identifies a focusing (de-focusing) medium being $n_2>0$ ($n_2<0$).
\\
I start considering paraxial linear (i.e. $\chi=0$)
X-wave solutions of (\ref{NLS}) 
(paraxial X-waves
 have been considered in detail in Ref. \onlinecite{Porras02})
, defined by
\begin{equation}
\label{Xeqsimple}
  \bigcirc\psi\equiv(\Delta_\perp -\partial_{tt})\psi=0\text{.}
\end{equation}
Introducing the complex variable $v=(\Delta-it)^2+r^2$,
with $\Delta$ a real valued arbitrary coefficient,
 I have from (\ref{Xeqsimple}) the equation 
\begin{equation}
6\partial_v \psi+4v  \partial_{vv} \psi=0\text{,}
\end{equation}
 from which 
$\psi=C_1/\sqrt{v}+C_2$. $C_1$ and $C_2$ are arbitrary complex
coefficients.
Note that both the real and the imaginary parts of this solution, are real-valued 
X-wave profiles. The former being the simplest X-wave, given by
(the branch cut for the square root is along the negative real axis)
\begin{equation}
\label{psiX}
\psi_X\equiv\Re(\frac{1}{\sqrt{v}})=\Re(\frac{1}{\sqrt{(\Delta-it)^2+r^2}})\text{.}
\end{equation}
Remarkably $\psi_X$ still holds 
when referring to the Helmotz equation, instead of the
 paraxial wave equation (see e.g. Ref. \onlinecite{Saari01}).
Its plot is given in figure \ref{figurepsiX}.

X-wave instability can be introduced in the same way as MI, i.e. by
perturbing the plane-wave solution of (\ref{NLS}):
$a=a_P\equiv a_0 \exp(ia_0^2 z)$, with $a_0$ a real-valued constant.
Letting $a=(a_0+\epsilon(x,y,t,z))\exp(i a_0^2 z)$ I have,
at first order in $\epsilon$:
\begin{equation}
\label{epsieq}
i\partial_z \epsilon+\bigcirc\epsilon+ \chi a_0^2 (\epsilon^*+\epsilon)\text{.}
\end{equation}
Writing
$\epsilon=\epsilon(r,t,z)=\psi_X(r,t)\mu(z)$ 
gives 
\begin{equation}
\label{mueqsimple}
i\partial_z\mu+ \chi a_0^2 (\mu+\mu^*)=0\text{.}
\end{equation}
The perturbation is thus
\begin{equation}
\label{epsilonsimple}
\epsilon=[\alpha+i(\beta+2 \chi a_0^2 \alpha z)]\psi_X(r,t)
\end{equation}
with $\alpha$ and $\beta$ arbitrary real-valued constants.
Equation (\ref{epsilonsimple})
 represents an X-wave which grows linearly along propagation (independently
on the sign of $\chi$), 
with amplification given by the intensity of the plane wave $a_P$.
This situation strongly resembles MI, where plane waves are exponentially
amplified at the expense of the pump beam, with a gain determined
by the pump intensity. For this reason, it is 
natural to refer to this process as {\it X-wave instability}.
As for MI, the amplified wave can be artificially
externally feeded, or it can be generated by noise.
\footnote{
It is remarkable that both X-waves and MI plane
waves, $\exp{(i\omega t+i k_x x+i k_y y)}$,
 fill all the space-time and have infinite energy.
Furthermore X-waves may be enriched by orthogonality properties 
as described in Ref. \onlinecite{Salo01b}
}
Note also that X-wave instability can be triggered by conical emission.
Indeed, as shown in \cite{Trillo02},
 the latter generates the required spectrum to
form an X-wave, which eventually gets amplified, as discussed above.

\begin{figure}
\includegraphics[height=40mm]{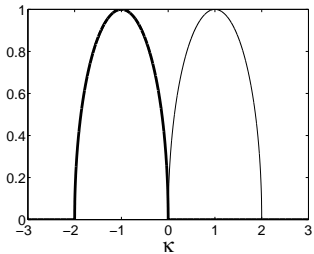}
\caption{\label{figuregain} Gain $\gamma$ Vs $\kappa$, the parameter identifying
the generalized X-wave, for focusing (thick line) and 
defocusing (thin line) media ($a_0=1$). }
\end{figure}

The previous treatment can be generalized in several ways. In 
particular it is possible to show that exponential amplification of 
X-wave-like beams can be attained. It is necessary
to enlarge the definition of X-waves, i.e. Eq. (\ref{Xeqsimple}), 
 by introducing the equation:
\begin{equation}
\label{genXeq}
  \bigcirc\psi=\kappa \psi\text{,}
\end{equation}
with $\kappa$ a real constant ($\kappa\neq0$ in the following).
Letting $\epsilon=\mu(z)\psi(r,t)+\nu(z)^*\psi(r,t)^*$ in (\ref{epsieq}),
and setting to zero the coefficients of $\psi$ and $\psi^*$ the following linear system is obtained:
\begin{equation}
\label{linsys}
\begin{array}{c}
i\partial_z \mu+\kappa \mu+a_0^2 \chi(\mu+\nu)=0\\
-i\partial_z \nu+\kappa \mu+a_0^2 \chi(\mu+\nu)=0\text{.}
\end{array}
\end{equation}
If $(\mu,\nu)=(\hat{\mu},\hat{\mu})\exp(\gamma z)$ I have
\begin{equation}
\left[ \begin{array}{cc}
i\gamma+\kappa+\chi a_0^2 & \chi a_0^2\\
\chi a_0^2 & -i\gamma+\kappa+\chi a_0^2 \end{array}
\right] 
\left[ \begin{array}{c} \hat{\mu}\\  \hat{\nu} \end{array} \right]=0\text{.}
\end{equation}
The
solvability condition yelds the allowed values for the gain 
$\gamma=\pm\sqrt{-\kappa(\kappa+2\chi a_0^2)}$.
The perturbation 
grows along $z$ if the following inequality
is satisfied: $-\kappa(\kappa+2\chi a_0^2)>0$.
Thus, for a focusing (de-focusing) medium, 
generalized X-waves with $-2 a_0^2<\kappa<0$ ($0<\kappa<2a_0^2$) are 
exponentially amplified.
The gain Vs $\kappa$, $\gamma=\sqrt{-\kappa(\kappa+2\chi a_0^2)}$, plotted 
in figure \ref{figuregain}, has a maximum value which corresponds
to the most exploding self-localized packet.
 
\begin{figure}
\includegraphics[width=60mm]{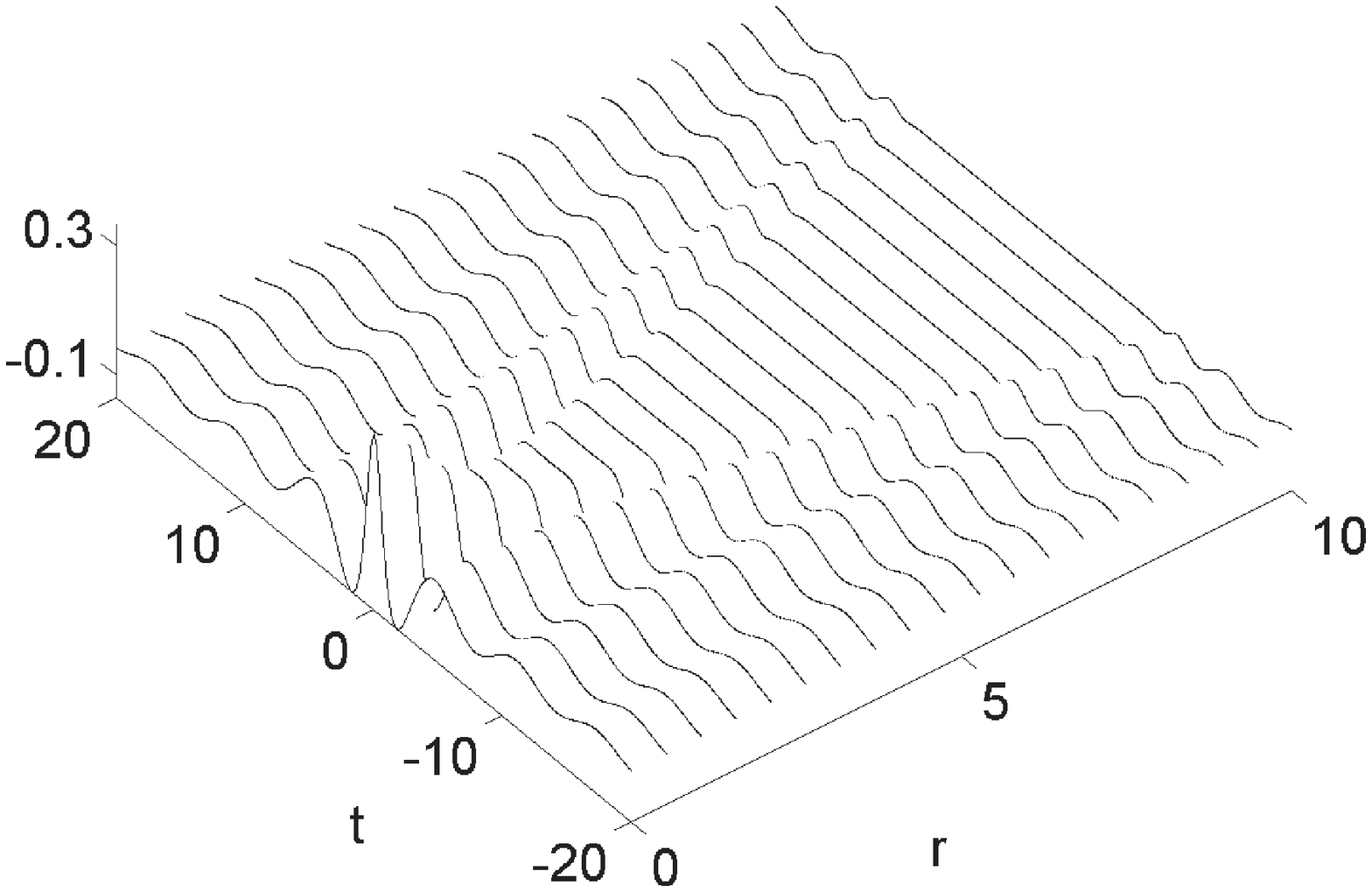}
\caption{\label{figurekpos} 3D plot of the generalized X-wave 
$\psi_\kappa$ when $\kappa=\Delta=1$.}
\end{figure}

\begin{figure}
\includegraphics[width=60mm]{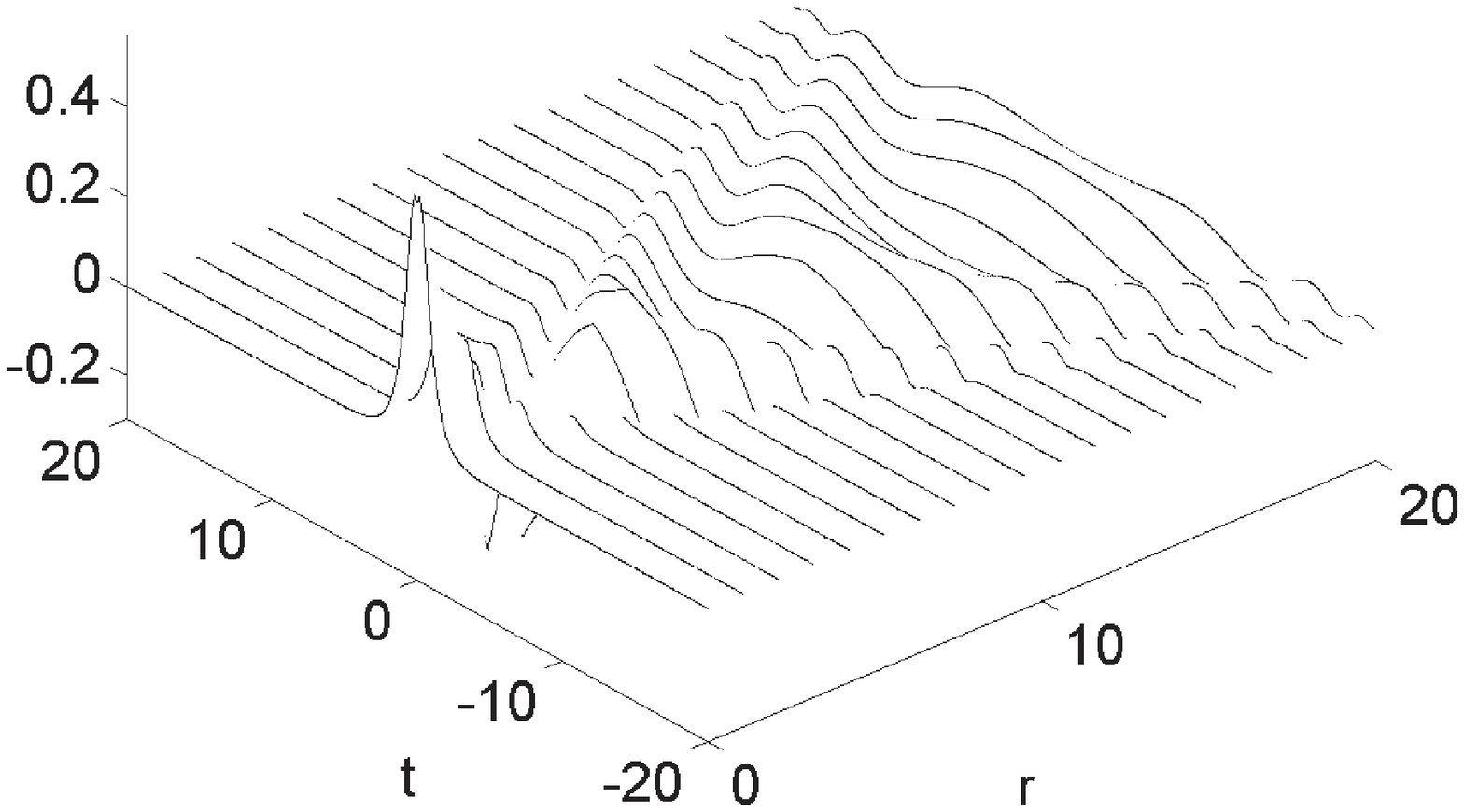}
\caption{\label{figurekneg} 3D plot of the generalized X-wave 
$\psi_\kappa$ when $\kappa=-\Delta=1$.}
\end{figure}

\begin{figure}
\includegraphics[width=60mm]{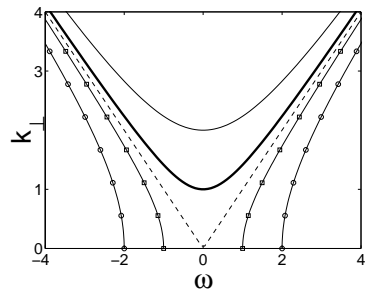}
\caption{\label{figurespectra}
Spatio-temporal spectra (transversal wavevector Vs angular spectrum)
for different generalized X-waves (thin line, $\kappa=-2$;
thick line, $\kappa=-1$; squares, $\kappa=1$; circles $\kappa=4$). 
The dashed line is the spectrum of
the simplest X-wave ($\kappa=0$).}
\end{figure}

Now the question arises as to whether or not Eq. (\ref{genXeq}) admits solutions resembling 
X-waves. Closed forms can be found proceeding as before: In term of $v$
Eq.(\ref{genXeq}) becomes: 
\begin{equation}
6\partial_v\psi+4 v\partial_{vv}\psi=\kappa\psi\text{,}
\end{equation}
whose general solution is 
\begin{equation} 
\psi=C_1 \frac{\exp(-\sqrt{\kappa v})}{\sqrt{v}}+
C_2 \frac{\exp(\sqrt{\kappa v})}{\sqrt{v}}\text{.}
\end{equation} 
A real valued, localised, generalised X-wave is given by
\begin{equation}
\label{psik}
\psi_\kappa=\Re(\frac{\exp(-\sqrt{\kappa v})}{\sqrt{v}})\text{.}
\end{equation}
(\ref{psik}) depends on two parameters $\Delta$ and $\kappa$;
while the first determines the decay constant as going far from the origin
in the $(r,t)$ plane, the latter completely changes the 
shape of the wave. 
Examples for $\Delta=1$ and $\kappa=1$ are shown
in figure \ref{figurekpos} and in figure
 \ref{figurekneg} for $\Delta=1$ and $\kappa=-1$. Note that they are similar to 
the Bessel pulse beams described in \cite{Sheppard01}.
To clarify the differences, I observe that the spatio-temporal spectrum
develops around the curve $\omega^2=k_\perp^2+\kappa$, with $\omega$ the angular frequency
corresponding to $t$ and $k_\perp$ the transversal wavenumber.
In figure \ref{figurespectra} I show the spectra for different $\kappa$,
with $\kappa=0$ corresponding to the simplest X-wave $\psi_X$.
The appearance of this lines in the spatio-temporal far field is a 
clear signature of the X-waves instability, and can 
be directly observed in the experiments. Note that
the spectra resemble the shape of the wave in the physical space 
as 
a consequence of their propagation invariance.

\begin{figure}
\includegraphics[height=35mm]{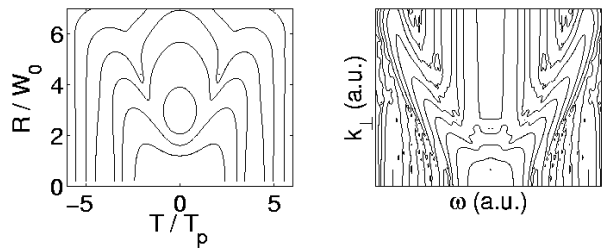}
\caption{
\label{figurenumerics}
Results of the numerical integration of Eq. (\ref{NLS})
after three diffraction lengths (
(Left) Level plots of $10\log(|u|^2)$;
(Right) Level plots of the square-modulus of the  spectrum in log scale of $u$.
}
\end{figure}

To show that X-wave instability can actually be observed in the 
experiments I numerically solved Eq. (\ref{NLS}).
I considered a gaussian input beam,
$A=A_0 \exp(-R^2/(2 W_0^2)-T^2/(2 T_p^2))$, whose intensity
 profile FWHM spot and duration are $70\mu m$ and $100 fs$.
Eq. (\ref{NLS}) is integrated with reference to fused silica, with $\lambda=800nm$
 ($n_0=1.5$,$n_2=2.5\times10^{-20}$ $k''=360\time10^{-28}$, SI units ),
peak power $P=1.5P_c$, being $P_c=(0.61 \lambda)^2\pi/(8 n_0 n_2)\cong2.6MW$ the 
critical power for self-focusing.
In figure \ref{figurenumerics} 
I show the spatio-temporal profile and the spectrum
(in log scale) after $3$ diffraction lengths $L_{df}=\pi n_0 W_0^2/\lambda$. 
Clearly an X-like profile is formed and the spectrum shows the features
 in Fig. \ref{figurespectra}. 

In conclusion I have shown that a plane wave in Kerr media
 gives rise to linear and exponential amplification of X-waves,
 thus leading to a significat beam reshaping.
A new class of X-waves is involved in this nonlinear process.
 The reported analysis provides insights
for the interpretation of the pulse-splitting of focused femtosecond beams,
and related phenomena, 
in the same way as  MI is relevant for solitons generation.
Indeed, the spostaneous formation of an X-wave can be another explanation
of the out of axis energy redistribution typically observed.
X-wave instability can also be an effective approach for the 
controlled generation of non-dispersive and non-diffractive pulses. 

These results appear to be susceptible of several generalizations,
as considering quadratric nonlinearity or vectorial effects,
and  have implications in all the fields encompassing
nonlinear wave propagation, as acustics, hydrodynamics and plasma physics. 
For example, X-wave instability (as well as nonlinear X-waves)
 is also expected in periodical Bose-Einstein
Condensates where Eq. (\ref{NLS})
 holds, being $t$ the direction of periodicity, in
the presence of negative effective mass. \cite{Konotop02,Kramer02}
\begin{acknowledgments}
I thanks S. Trillo and E. Recami for fruitful discussions,
and the Fondazione Tronchetti Provera for the financial support. 
\end{acknowledgments}

%
\end{document}